# THE UFFO (ULTRA-FAST FLASH OBSERVATORY) PATHFINDER

I.H. Park[1,2], B. Grossan[3,4], H. Lim[2], J.W. Nam[1], Pisin Chen[5,6], B. A. Khrenov[7], Y.-K. Kim[8], C.-H. Lee[1,9], J. Lee[1], E. V. Linder[2,3,4], M. Panasyuk[7], J.H. Park[1,10], G. F. Smoot[2,3], Z. L. Uhm[1,2]

for the UFFO Collaboration[†]

[1]Research Center of MEMS Space Telescope (RCMST), Ewha Womans University, Seoul 120-750, Korea
[2]Institute for Early Universe (IEU), Ewha Womans University, Seoul 120-750, Korea
[3]Berkeley Center for Cosmological Physics (BCCP), University of California, Berkeley, California 94720, USA
[4]Space Sciences Laboratory, University of California, Berkeley, California 94720, USA
[5]Leung Center for Cosmology and Particle Astrophysics & Department of Physics and Graduate Institute of Astrophysics, National Taiwan University, Taipei, Taiwan 10617
[6]Kavli Institute of Particle Astrophysics and Cosmology, Stanford Linear Accelerator Center, Stanford University, Stanford, CA 94305, USA
[7]D.V. Skobeltsyn Institute of Nuclear Physics (SINP), Moscow State University, Moscow 119992, Russia
[8]School of Electrical Engineering and Computer Science, Seoul National University, Seoul 151-600, Korea
[9]Department of Physics, Pusan National University, Pusan 609-735, Korea
[10]Electronics and Electrical Engineering, Dankook University, Kyungkee 448-701, Korea

## ABSTRACT

Hundreds of gamma-ray burst (GRB) UV-optical light curves have been measured since the discovery of optical afterglows, however, even after nearly 5 years of operation of the SWIFT observatory, only a handful of measurements have been made soon (within a minute) after the gamma ray signal. This lack of early observations fails to address burst physics at the short time scales associated with burst events and progenitors. Because of this lack of sub-minute data, the characteristics of the UV-optical light curve of short-hard type GRB and rapid-rising GRB, which may account for ~30% of all GRB, remain practically unknown. We have developed methods for reaching the sub-minute and the sub-second timescales in a small spacecraft observatory appropriate for launch on a microsatellite. Rather than slewing the entire spacecraft to aim the UV-optical instrument at the GRB position, we use rapidly moving mirrors to redirect our optical beam. Our collaboration has produced a unique MEMS (microelectromechanical systems) micromirror array which can point and settle on target in only 1 ms. This technology is proven, flying successfully as the MTEL (MEMS Telescope for Extreme Lightning) on the Tatiana-2 Spacecraft in September 2009 and as the KAMTEL on the International Space Station in April 2008. The sub-minute measurements of the UV-optical emission of dozens of GRB each year will result in a more rigorous test of current internal shock models, probe the extremes of bulk Lorentz factors, and provide the first early and detailed measurements of fast-rise and short type GRB UV-optical light curves.

## 1. INTRODUCTION

### 1.1 GRB Science

The discovery of optical afterglows of gamma-ray bursts (GRB) was a monumental event in modern astrophysics, ending the thirty-year mystery of the GRB distance scale. The study of GRB UV-optical afterglows and their host galaxies has led to knowledge of the origin of some types of GRB and the discovery of the most distant object known (GRB090423 at $z$=8.2). Much progress has been made in GRB science since the launch of the SWIFT observatory (Gehrels et al. 2004) nearly 5 years ago. The observations from SWIFT did not produce a simple picture of

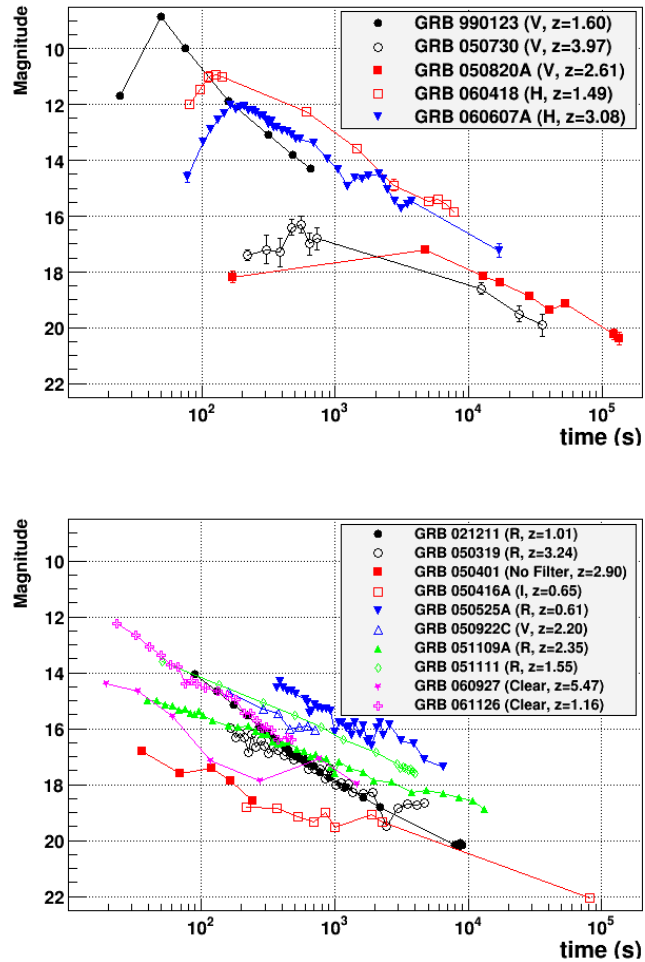

**Figure 1. Variation in GRB Optical Light Curve Rise Time and Shape** (from the Panaitescu & Vestrand 2008 sample). The top plot shows the light curves of the Fast-Rising GRB classes identified in Panitescu & Vestrand (2008); the light curves of the Decay class are shown in the lower figure. Note that the fastest-rising light curves are poorly sampled before their peak. This survey of rapid response bursts shows that the number of bursts with measured rise times falls off rapidly below ~100 s. The measured (i.e. slow or fast rise) bursts exhibit a correlation of rise time and luminosity (Figure 2); since the rise time is not known for the Decay class bursts, the correlation cannot be tested among all these bursts.

GRB, but rather documented the richness and complexity of this phenomenon. Just a few years ago, GRBs were believed to be of only two types, distinguished only by their gamma-X emission: a shorter, hard spectrum burst, with duration of gamma-X emission less than two seconds, and a longer, soft type burst. After some 370 UV-optical observations by UVOT (the UV-optical telescope on SWIFT) made to date, a huge variation in light curves has been observed, especially in the early rise time. Figure 1 shows a sample of the light curve measurements made as soon as possible after GRB triggers. There appear to be distinct classes of fast-rising ($t_{peak} < 10^2$ s) and slow-rising bursts (Panaitescu & Vestrand 2008). Additionally, the light curves are complex, with decays, plateaus, changes in slope, and other features that are not yet understood. Panaitescu & Vestrand (2008) claim that the luminosity distribution of the fast-rising bursts at ~ $10^3$ s is quite narrow, and has promise as a kind of "standard candle" which would make GRBs useful as a cosmological probe of the very high red shift universe. In order to move this possible trend to the status of a refined tool, a larger sample of such objects is required, and in particular, better resolution is required at early times: The fastest bursts often have just one measurement in their rising phase – hardly enough to understand the physics in this regime – and many other bursts have no early measurements at all. Only 6 GRB in this study were measured at less than 100 seconds after their burst trigger and *not a single measurement* at less than 15 s after trigger. Are there more features in the early light curve that are missed by such sparse sampling? Does any feature of the rise correlate with the luminosity or a particular aspect of the physics? How many bursts are mis-classified because the rapid rise was missed? The need for earlier measurements (faster UV-optical response after the initial gamma-ray burst) is clear and compelling.

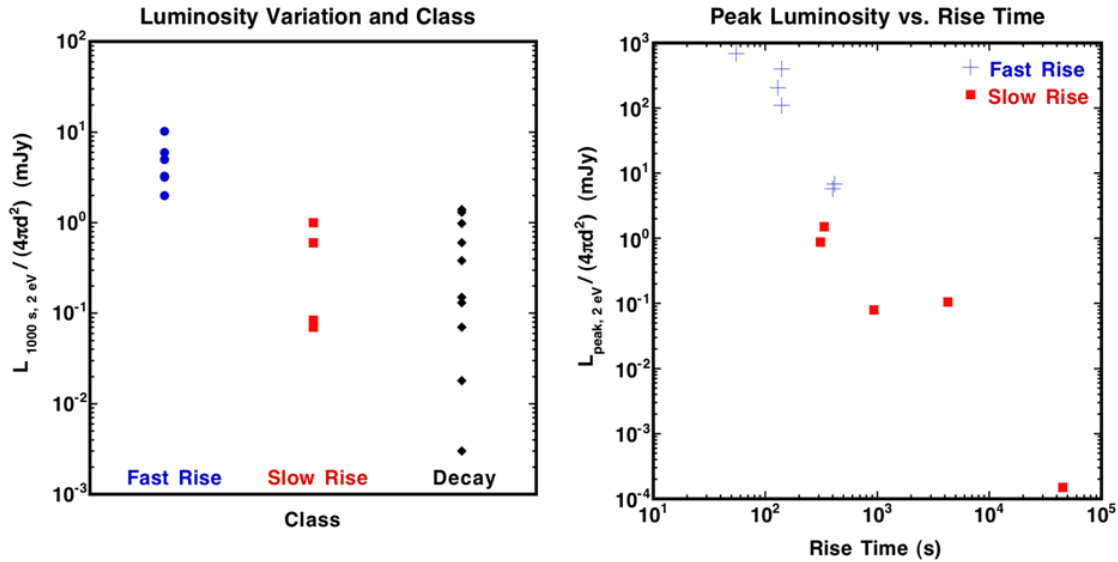

**Figure 2. Panaitescu & Vestrand Luminosity Correlation**. The left plot shows the flux for each GRB, transformed to a uniform band emitted frame at *z*=2, corrected for extinction and extrapolated to t=1000 s, taken directly from Panaitescu & Vestrand (2008). This "standardized flux", which is proportional to luminosity, varies by less than one dex among the fast-rise GRB. The slow-rise GRB vary by about 1.5 dex, and the decay GRB vary much more. The right figure shows the correlation of luminosity and peak time claimed by Panaitescu & Vestrand (2008), suggesting that GRB could be "calibratable" standard candles. Confirmation requires more measurements before peak, i.e. at very early times.

The recent progress in short-hard GRBs is extremely exciting. As of this writing, only about 8 short-hard GRBs had UV-optical measurements, again suffering from poor time resolution in their light curves. What is the shape of the rise? Is the shape homogeneous? The physical origin of this type of burst remains an outstanding mystery, so any hints as to this origin would be extremely valuable. Because of the short time scale for the gamma-X light curves, and the lower bolometric luminosity, these bursts are believed to originate from the merger of compact objects. Is there any prompt UV-optical emission from such events? What would we see if we observed more of these events in the sub-minute or sub-second regime? Are there ultra-short events on the accretion disk dynamical timescale of compact objects (that are beamed so we can see them)? Earlier observations would answer these questions and open a new window probing compact object structure, populations, and evolution.

"Dark" GRB are those that stand out as having a very faint optical compared to X-ray afterglow. Only recently, extinction has been found to be the dominant source of dark GRB (Perley, D. A. et al., 2009). An alternative scenario, however, suggests that some "Dark" GRB are simply due to a faster decay for optical than X-ray emission. In this scenario, the optical emission fades in less than ~ $10^2$ s, so that most observations would not detect the optical afterglow. Better short-timescale observations would shed light on this two-mechanism model.

Measurement of early UV-optical emission can serve as a probe of the physical conditions in the GRB fireball at short times. A simple, nearly model-independent argument (Molinari et al. 2007), shows that the bulk Lorentz factor depends on the time of the early UV-optical emission peak (see Figure 3). Measurement of the peak will therefore provide a measurement of the bulk Lorentz factor.

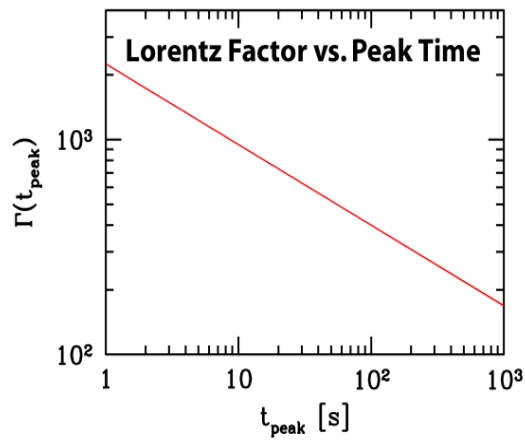

**Figure 3.** Using the method of Molinari et al. 2007, the bulk Lorentz factor Γ may be measured via the peak emission time. Measuring in the sub-minute regime will probe the limits of bulk Lorentz factors, constraining central engine models.

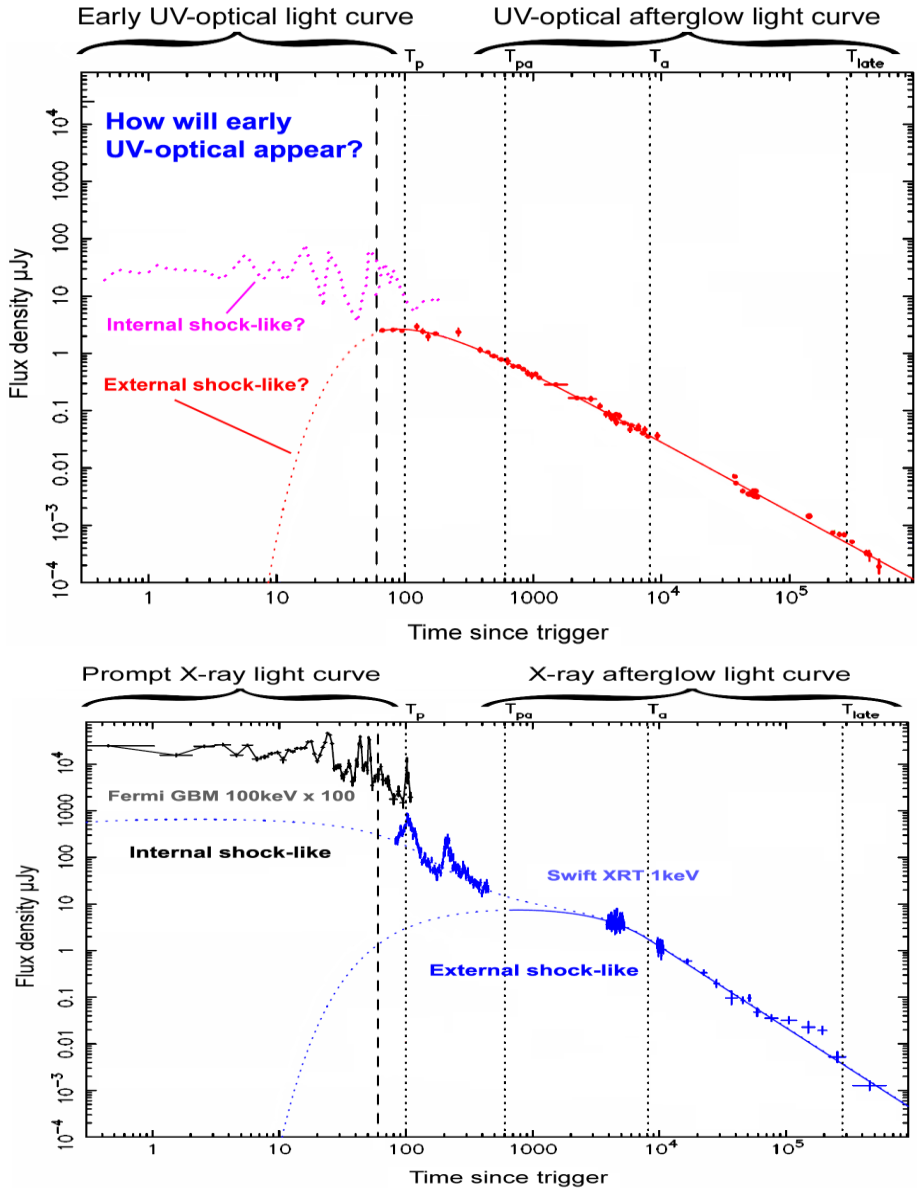

**Figure 4. Internal vs. External Shock Signatures.** The prompt X-ray light curve (bottom) is jagged, consistent with internal shocks. The late time afterglow X-ray and UV-optical light curves (top) are smooth, consistent with external shocks. Does the early UV-optical light curve show "internal shock-like" – jagged, or "external shock-like", smooth behavior? Observation of this behavior would show the role of internal shocks at early times. Figures include work from Page et al (2009), used by permission.

Currently, UV-optical emission at early times in typical bursts is believed to come from external shocks, and predicted to have a smooth, monotonic rise (e.g. see Piran 2004 and references therein). Observation of an early time UV-optical light curve that more closely resembles a gamma-X light curve, jagged, and with multiple peaks, would clearly indicate the presence of internal shock produced prompt emission in this band (see Figure 4). Sub-minute measurements would be required to learn more about such prompt emission. What are the prospects for pushing to shorter time scales in GRB measurements?

## 1.2  Current Limits of Rapid Response Measurements

The SWIFT observatory produces UV-optical light curves by first serendipitously detecting the onset of a GRB within the very large field of the Burst Alert Telescope (BAT; Barthelmy 2004). The BAT then produces a crude sky position via a standard coded mask technique. After this, the entire observatory spacecraft slews to point the UVOT (and other instruments) at the GRB position. After slewing, a period of time is required for the pointing to stabilize, after which a series of UVOT exposures begins. This system has produced numerous detections of optical afterglows associated with GRB, a spectacular success by any measure. However, after nearly 5 years of operation, only a handful of responses have occurred in less than 60 s. Figure 5 shows that the frequency of responses falls off for response times below 100 s, with an almost complete cutoff by 60 s. Due to finite mission lifetime, SWIFT cannot be expected to significantly increase this number of sub-minute responses.

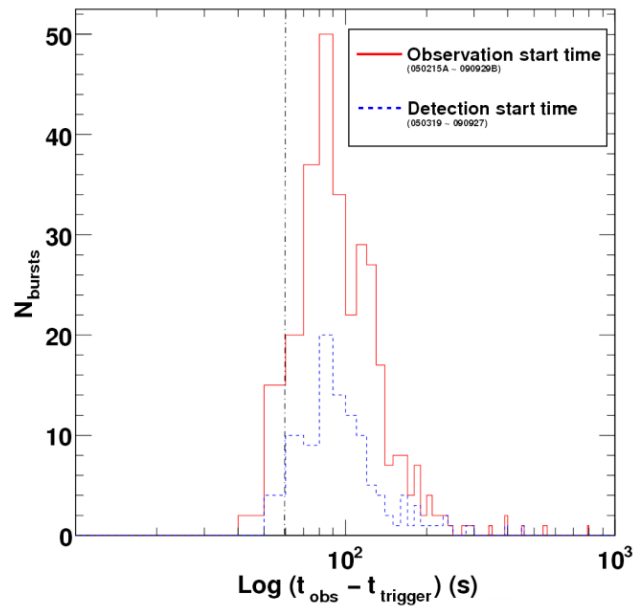

**Figure 5.** Distribution of SWIFT UVOT response times. Only a small number of responses have occurred below 60 s (vertical dash-dot line), and none at all below 40 s, with no detections below 50 s. UFFO will explore the blank parameter space on the left side of the figure, the fast- and ultra-fast regimes below 40 s, in a systematic survey, for the first time

A handful of ground-based rapid-response instruments such as ROTSE, TORTORA, and Super-LOTIS have response times as short as ~20 s, but due to their small size, and to the limitations of ground-based observing, including daytime and weather, together these instruments have managed only a handful of rapid detections (e.g. Akerlof et al. 1999). Because of the ability of space-based telescopes to detect photons throughout the UV-optical band without scattering or absorption, the 30 cm aperture UVOT telescope compares favorably in sensitivity to a 4-m ground-based telescope (Roming et al. 2005), and such telescopes are not capable of sub-100 s response. The SWIFT limit of 60 s response is therefore the practical minimum for sensitive UV-optical GRB studies for the near to mid-term future.
.

## 2. A NEW APPROACH: MOVE THE OPTICAL PATH, NOT THE SPACECRAFT

The time to rotate the spacecraft to slew the UV-optical telescope is the limiting factor in the SWIFT response time; not only must the entire spacecraft mass be rotationally accelerated and decelerated, but after the movement some additional time is required for any vibrations to cease. Vibration can cause smearing of the telescope point spread function (PSF) and loss of sensitivity. Our approach to reduce this time is to move only small and light mirrors to redirect the optical path, instead of the entire spacecraft. We find that various types of rotating mirrors move across our entire field of view (FOV), point, and settle in less than 1 s.

What about the sub-second response regime? Our lab consortium has produced small mirror arrays driven by MEMS devices. These MEMS mirror arrays (MMA), fabricated like other microelectronics devices, are very small and light and can move, point, and settle in less than a few ms. This project is the ideal opportunity to apply this new technology to transient and GRB science, making GRB measurements for the first time on timescales as short as milliseconds after trigger and target location.

We have designed a small telescope and X-ray trigger system to provide imaging measurements separately in the "fast" sub-minute, and in the "ultra-fast" sub-second regime, beginning only ~ 1 ms after trigger+target location. The system was designed to (i) fit the constraints of the Lomonosov spacecraft, (ii) use all pre-proven technologies and (iii) to be available for fast delivery. In order to use a proven successful design and approach, we essentially scaled down the SWIFT UVOT system to fit the available mass and size requirements, but added a rotating mirror and MMA. We also take triggers from a wide-field coded mask camera. We take the main constraints for inclusion in Lomonosov to be less than 60 kg total instrument mass, and 800 cm maximum length. In the remainder of this proposal, we give details on a complete instrument to accomplish this task for the first time.

## 3. SLEWING MIRROR TELESCOPE (SMT)

### 3.1 Concept

The key idea of the SMT UV-optical telescope is very fast pointing of the narrow-FOV UV/optical telescope using a fast rotatable mirror plate covered by micromirror devices (Park 2004). A schematic of this concept is shown in Figure 6 (left). The parallel rays are directed on-axis with respect to the fixed optics by the rotatable mirror system. The net effect is to steer the UV-optical instrument beam, instead of moving the telescope or the spacecraft itself. The beam can be steered by two-axis rotation of the mirror plate or rotation of the individual micromirror devices. In either case, the mirror rotation angle is ±15° off axis, resulting in an accessible FOV of 60° x 60°, without the aberration inherent in wide-field optical systems.

Our simulations of our segmented MMA show that the SMT PSF will have a FWHM of about 1 arcsec with the micromirrors at zero tilt. When the micromirrors are tilted, however, the PSF spreads to a FWHM of 2 arcmin due to the difference in beam path length created by tilt of elements. Therefore, we use the MMA to steer the beam to measure very early photons, starting $\sim 10^{-3}$ s after trigger+location; we perform high-resolution imaging of the source later, using the mirror plate to steer the beam (with the micromirrors at zero tilt), because this motion takes much longer, $\sim 1$ s.

**Table 1.** SMT Specifications

| | |
|---:|:---|
| Telescope | Ritchey-Chrétien + MMA with motorized plate |
| Aperture | 20 cm diameter |
| F-number | 6 |
| Detector and / Operation | Intensified CCD with MCP / Photon Counting |
| Field of View | 17 x 17 arcmin |
| Detection Element | 256 x 256 pixels |
| Telescope PSF | 1 arcsec @ 350 nm |
| Pixel Scale | 4 arcsec |
| Location Accuracy | 0.5 arcsec |
| Wavelength Range | 200 nm – 650 nm |
| Color filters | 8 |
| Sensitivity | B=23.5 in white light in 1000 s (16 in 1 s) |
| Bright Limit | mv = 6 mag |
| Data taking start time after trigger+location | 2 ms (as fastest) |
| Data Rate | 1 Gbytes/day |
| Mass, Power consumption, Size | 20 kg, 20W, 67.5 cm x 68 cm x 29 cm |

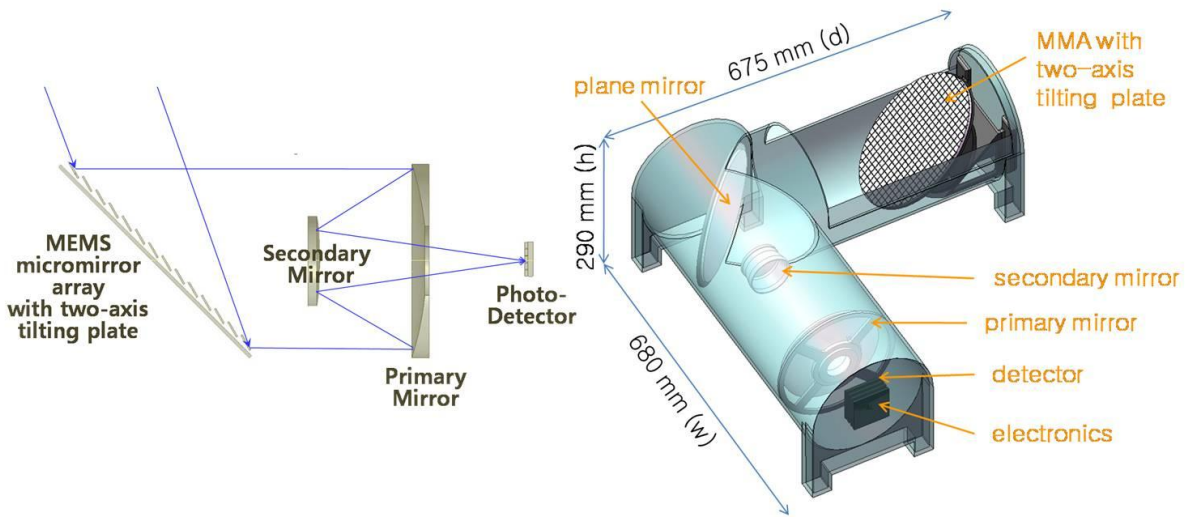

**Figure 6**. **Design of the SMT.** A simplified schematic of the optical design showing the location of the MMA and rotating plate is given at left; a more detailed drawing of the flight

### 3.2 Telescope Optical Layout & Design

We constrained the telescope to fit within 80 cm in its largest dimension, and to have a 20 cm aperture for sensitivity comparable to that of SWIFT UVOT. We used a similar Ritchey-Chrétien design, with the details given in Table 1, and shown in Figure 6.

### 3.3 MEMS Mirror Array (MMA)

Each micromirror in our array devices is driven by two-axis electrostatic vertical comb actuators that allow continuous changes to the tilt angle of the mirror plate around two orthogonal axes. Figure 7 shows a schematic view of a single micromirror element and SEM images of our 8 × 8 micromirror array (Park et al. 2008, Kim et al. 2009).

A small 3 mm x 3 mm prototype device was used on an earlier telescope, and delivered to the International Space Station (ISS) on April 11-17, 2008. The larger 8 × 8 micromirror array, each reflector unit 340 µm × 340 µm, was carried into orbit on a Russian microsatellite, Tatiana-2, on September 17, 2009 (Nam et al. 2008, Park et al. 2008). The primary aim of this mission is to observe Transient Luminous Events (TLEs) in the upper atmosphere over mission duration of at least one year. The micromirrors have been space-qualified for operation on the ISS and in earth orbit. The array functioned successfully on the ISS, and its performance is described in Yoo et al. (2009). So far, all systems are working well on the instrument on board Tatiana-2.

Production of an MMA large enough to fill the entire aperture of the SMT is a significant technological challenge. The array will be composed of 300 × 300 reflector pixels, each 1 mm × 1 mm, meeting the same requirements as previous MMA, ±15° tilt, ≤ 2 arcmin PSF spread. We expect to have the first production of these devices for the UFFO in early 2010.

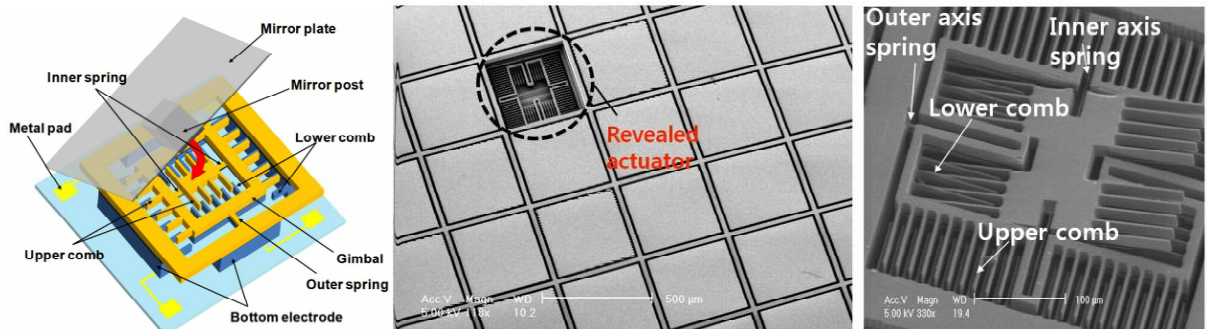

**Figure 7.** MEMS Mirror Arrays. A schematic view of the micromirror unit cell is shown at left, followed by the scanning electron microscope images of a fabricated micromirror array at center, and a single actuator, right.

### 3.4 Fast/Ultra-Fast Operation

Observations with the UFFO proceed as follows:

(0) Standby Mode: $t_{resp}$ = - infinity to 0. The mirror plate and MMA are pointing at the center of the field.

(1) Ultra-fast Slew: $t_{resp}$ = 0 to ~10 ms. On receipt of trigger+location, the MMA is directed to point at the target.

(2) Ultra-fast Data Acquisition: $t_{resp}$ = 0.01 to 1 s. The MMA tracks the target for one second, and data is recorded at maximum rate.

(3) Fast Slew : $t_{resp}$ = 1 to ~ 2 s. Mirror plate tilts to point at target.

(4) Fast Data Acquisition: $t_{resp}$ = 2+ s. Data are acquired with MMA at zero deflection; source is tracked by mirror plate tilt.

In ultra-fast mode, the mirror plate remains aimed at the middle of the field while the MMA elements point at the target. In fast mode, the MMA elements are in standby or zero deflection mode, and the rotatable plate aims at the source. The first mode minimizes response time, the second maximizes sensitivity. The MMA provides ultra-fast response times, but at some cost in terms of PSF broadening, as described above. There is essentially no degradation in performance for the MMA in standby mode, and the full resolution (and therefore sensitivity) of the telescope is obtained in fast mode.

### 3.5   Detector System

We choose a detector chain nearly identical to that of UVOT by using the same UV sensitive intensified CCD system with microchannel plate (MCP). The size of MCP microchannel and the UV sky background essentially determine our pixel size and FOV to be the same as UVOT. We designed our gamma-X trigger which localizes the target within the 17 arcmin FOV of the SMT at our threshold fluence.

## 4.   UFFO BURST ALERT & TRIGGER TELESCOPE (UBAT)

### 4.1   Instrument Design

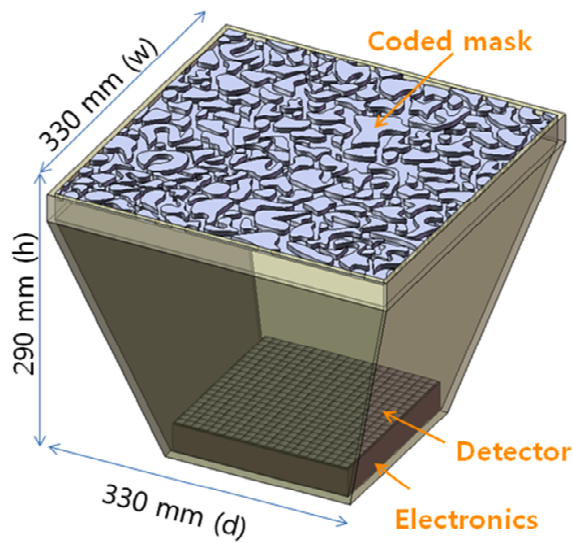

**Figure 8.** Structural design of UBAT.

The UFFO Burst Alert & Trigger telescope (UBAT) will be similar to the SWIFT BAT X-gamma trigger camera, using a coded mask aperture camera scheme for good position detection for transients. However, in order to get response over a wider energy range, making the camera more sensitive to broad-band hard sources including GRB, a design including CdTe detectors is used, resulting in a sensitive energy range of 4 - 250 keV. This is similar to the Eclair/CXT design (Schanne 2008). In order to fit the UBAT into the Lomonosov constraints, dominated by the mass constraint of ~ 20 kg, we use a detection area of 1024 $cm^2$. The resulting sensitivity is 310 mCrab in 10 s at 5 $\sigma$. The specifications of UBAT are given in Table 2 (see also Figure 8).

## 4.2 Operation

We plan to trigger SMT observations on detecions of transient sources matching the general characteristics of GRB sources. We plan to allow triggers for secondary science targets (e.g. extreme activity of galactic or halo compact objects) or triggers of low significance and/or GRB identification figures of merit, but we allow for interruption of low-priority events with higher priority events, and erasure of low-priority event data in case of memory limitations. Fortunately, the SWIFT BAT development and experience has led to an extensive body of work (Barthelmy 2004) on trigger methods. The primary differences with our system are that since we will not operate in a pointing mode, we will need to add "software tracking" to compensate for spacecraft motion, and we will need to switch between threshold settings for approach of the galactic plane where transients are common but where GRB UV-optical emission may not be observed due to galactic dust obscuration.

**Table 2.** Parameters of UBAT

|  |  |  |
|---|---|---|
| Overall | Mass of the camera | 20 kg |
|  | Energy range | 4 – 250 keV |
|  | Telescope PSF | ≤ 17 arcmin |
|  | Source position accuracy | ≤ 10 arcmin > 7$\sigma$ |
|  | Field of view | ~2 sr (89˚ x 89˚) |
|  | GRB detection rate | 67/yr. |
| Detector plane | Compounds | CdTe |
|  | Effective area | 307 $cm^2$ |
|  | Pixel size | 1.2 x 1.2 x 1 $mm^3$ |
|  | Number of pixels | 146 x 146 (21316) |
|  | Spectral energy Resolution | ≤ 2 keV FWHM at 60 keV |
|  | Sensitivity | 310 mCrab for a 10 exposure at 5.5$\sigma$ 4-50keV |
| Passive shielding | Compounds (out to in) | Pb, Cu, Al |
|  | Absorption @ 4-50 keV | 100 % |
| Coded mask | Compounds | Ta-W alloy |
|  | Total size | 1024 $cm^2$ |
|  | Mask to detector plane distance | 25 cm |

### 4.3 Source Location

We designed the FOV of the BAT to match the large FOV accessible by the mirrors of the SMT, while at the same time providing GRB localizations within the field size of the SMT detectors. Reducing the mask-detector separation appropriately (see Caroli et al. 1987), the instrument will have a location accuracy of 10 arcmin over its FOV for 7 $\sigma$ sources. Source location is achieved by acquiring detector images and then using convolution operations with these images and the mask pattern. In SWIFT, this calculation process can take up to seven seconds. We plan to implement this algorithm in much faster and more modern FPGA and microprocessors. Based on our experience with FPGA programming and image processing, we estimate a few hundred ms for this task.

## 5. UFFO Payload Parameters

The design for the UFFO payload is presented in Figure 9. The TUS instrument for ultra high energy cosmic ray measurement will also fly on the Lomonosov spacecraft. The TUS looks down, and the UFFO will be on the upper side of the spacecraft looking up. The characteristics of the UFFO instrument package are given in Table 3. Because our mirrors will track sources, we have very modest requirements for spacecraft stability.

Table 3. Parameters of spacecraft and payload

| | |
|---:|:---|
| Name of Spacecraft | Lomonosov |
| Platform Provider | VNIIEM |
| Launch Date Expected | Nov. 2011 |
| Orbit Height | 500 ~ 600 km |
| Total Mass | 300 ~ 800 kg |
| Mission Lifetime | 3 years |
| Payload | TUS for observation of Ultra High Energy Cosmic Ray and UFFO Pathfinder for GRBs |
| Total Payload Mass | 100 ~ 150 kg (TUS: 80 kg, UFFO: 30 ~ 60 kg) |
| Total Payload Power | 100 ~ 150 W |

## 6. EXECUTION PLAN

Our lab has now had two successful flights of MEMS space telescopes, with flawless operation of MMA, optics, electronics and communications equipment. The experience has provided us with valuable experience in construction of optics and electronic systems for the space environment.

We expect to have our new larger-scale MMA (1 cm x 1 cm) by the end of 2010. Our goal is to produce and test the full field-of-view covering MMA mosaic by mid-2011. We will test the large mosaic for pointing uniformity and speed.

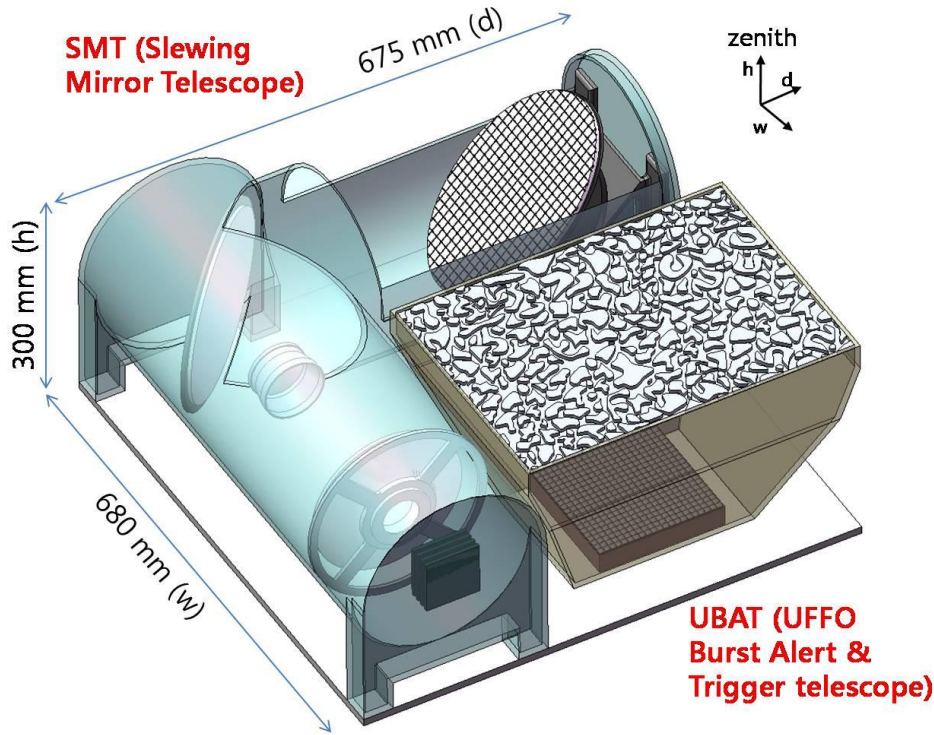

**Figure 9.** UFFO Pathfinder Payload

Coded mask aperture telescopes have an extensive heritage and are well understood and documented. Our lab has extensive experience providing high-energy physics instruments and electronics. However, we plan to have collaboration with some colleagues with direct and extensive experience with these types of X-ray detectors. We are currently in negotiations with several experienced groups to collaborate on the design optimization for Lomonosov and the fabrication of the UBAT camera. No technology needs to be developed for this instrument. The only development required is to optimize and adapt our coded mask camera to Lomonosov.

## 7. EXPECTED RESULTS & IMPACT

In order to estimate our event rate, we examined the fluence distribution of SWIFT BAT GRB which triggered SWIFT UVOT observations during the first 5 years of SWIFT operation. (See Figure 10. This is a more conservative number than the total rate of SWIFT BAT GRB). Scaling by our estimated sensitivity to that of SWIFT BAT, we find that we will still receive an expected 69 GRB triggers for SMT per year from UBAT. Of these, we expect 2.4 short-hard triggers per year. The actual number of SMT observations that we accomplish should cover about the same number, unless our orbit has significantly more restrictions than that of SWIFT. It is likely that some reduction in these numbers could result from our inability to point away from the galactic plane, unlike SWIFT. Because short bursts are hard, and because our detectors have more low-energy response than BAT, some reduction in the short-hard rate may result. (Normally when predicting event rate, one considers the "roll-off" of detection efficiency at low flux. Here, however, we consider only events brighter than 8σ, so the roll-off due to detection efficiency is small. The obvious roll-off of the SWIFT distribution in the figure is dominated by the well-known finite space distribution of GRB, not the detection efficiency roll-off; e.g. Fishman 1994.)

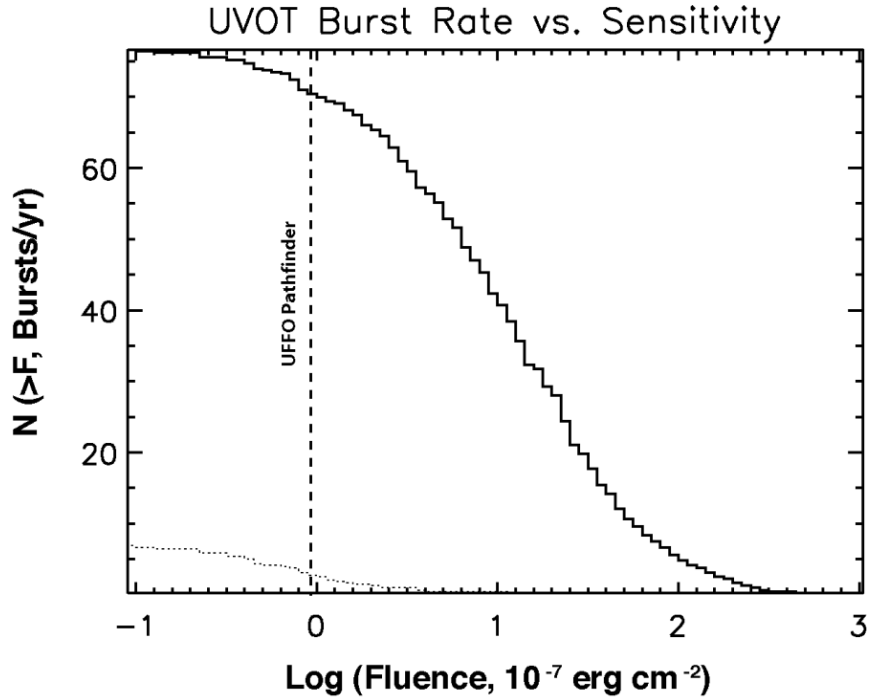

**Figure 10**. Cumulative SWIFT Fluence Distribution for UVOT Triggers. The histogram shows the integrated or cumulative number of GRB per year triggering UVOT observations as a function of gamma-X fluence for the period 2005 Jan. 24 through 2009 Sep. 21. The number of all bursts are given by solid line and short hard bursts by dotted line. Our UBAT sensitivity is ~10 times poorer than SWIFT BAT. However, the fraction of bursts actually lost is remarkably small (assuming no change in trigger efficiency between the two different bands, and scaling fluence sensitivity as flux sensitivity).

Barring significant malfunctions, UFFO will provide sub-minute UV-optical measurements for dozens of GRB within the first year of operation. These measurements will be the first ever under 20 s after the gamma ray signal, and will make up the first ever large-sample, systematic survey of emission in the sub-minute regime.

We note that there is an exciting synergy with the TUS instrument. In the case that a GRB produces neutrino or other cosmic ray signals, extensive air showers are expected. Our experiment would detect the source GRB event in X-gamma and UV-optical photons, while the TUS instrument would detect the particle shower, measuring the source position and arrival times of the particles with accuracy of up to 1° and 10 μs, respectively. The neutrinos with the long horizontal showers can be measured with 0.1° accuracy. Such measurements could enable the first measurements of neutrino masses from GRB emission, and would serve as an exciting new measure of photon and particle dispersion relations, of great interest to fundamental particle physics, cosmology and relativity tests.

## 8. SUMMARY

We have proposed the UFFO instrument, a 20 cm UV-optical telescope and X-gamma camera for observations of early emission from GRBs, to be launched on the Lomonosov satellite. In the ultra-fast mode, our new MMA-based response will produce measurements beginning within milliseconds after receipt of trigger+location.

The SMT UV-optical instrument will, for the first time, measure the first ~ 500 ms to 60 s of UV-optical emission after the onset of GRB emission and localization. We expect to observe ~ 70 GRBs/yr., and of those, roughly 2.4 short-hard type bursts/yr. are expected. These measurements will, for the first time, provide well-sampled UV-optical measurements in the sub-minute regime.

Such measurements will permit measuring or constraining internal shock emission starting from soon after the burst. They will also permit measurement of the bulk Lorentz factor in a much larger fraction of GRB than before, constraining models of GRB central engines. For the first time, the rise of short-hard GRBs will be clearly time-resolved as either measurements or upper limits, providing important hints as to the origin of these still-mysterious events. Finally, the large number of GRBs observed assures the characterization of a significant number of fast-rise class GRBs, properly time-resolved during the rise phase, an essential step in testing GRB as cosmological tools.

**† The UFFO Collaboration**

(Contact: ipark@ewha.ac.kr, Bruce_Grossan@lbl.gov)

J.A. Jeon[1], A.R. Jung[1], S.M. Jeong[1], J.E. Kim[1], H.Y. Lee[1], J. Lee[1], G.W. Na[1], J.W. Nam[1], S. Nam[1], I.H. Park[1,6], J.E. Suh[1]
[1]*Research Center of MEMS Space Telescope (RCMST), Ewha Womans University, Seoul 120-750, Korea*

B. Grossan[2,3], E. V. Linder[2,3,6], G. F. Smoot[2,6]
[2]*Berkeley Center for Cosmological Physics (BCCP), University of California, Berkeley, California 94720, USA*
[3]*Space Sciences Laboratory, University of California, Berkeley, California 94720, USA*

B.A. Khrenov[4], G.K. Garipov[4], M. Panasyuk[4], P. Klimov[4]
[4]*D.V. Skobeltsyn Institute of Nuclear Physics (SINP), Moscow State University, Moscow 119992, Russia*

C.-H. Lee[1,5]
[5]*Department of Physics, Pusan National University, Pusan 609-735, Korea*

H. Lim[6], Z. L. Uhm[1,6], J. Yang[6]
[6]*Institute for Early Universe (IEU), Ewha Womans University, Seoul 120-750, Korea*

J.H. Park[1,7]
[7]*Electronics and Electrical Engineering, Dankook University, Kyungkee 448-701, Korea*

J.Y. Jin[1,8], Y.-K. Kim[8], B.W. Yoo[1,8]
[8]*School of Electrical Engineering and Computer Science, Seoul National University, Seoul 151-600, Korea*

Y.-S. Park[1,9], H.J. Yu[1,9]
[9]*School of Physics and Astronomy, Seoul National University, Seoul 151-742, Korea*

S.W. Kim[10]
[10]*Korean Astronomy and Space Science Institute, Daejeon 200-200, Korea*

Y.-Y. Keum[1,11]
[11]*Department of Physics, Korea University, Seoul 136-701, Korea*

Pisin Chen[12,13], C.-K. Huang[12], M. Huang[12]
[12]*Leung Center for Cosmology and Particle Astrophysics & Department of Physics and Graduate Institute of Astrophysics, National Taiwan University, Taipei, Taiwan 10617*
[13]*Kavli Institute of Particle Astrophysics and Cosmology, Stanford Linear Accelerator Center, Stanford University, Stanford, CA 94305, USA*

Y.-H. Chang[14]
[14]*Department of Physics, National Central University, Jung-Li 32001, Taiwan*

M.-H. Huang[15]
[15]*Department of Energy and Natural Resource, National United University, Miao-Li 36003, Taiwan*